\documentclass[prb,twocolumn,superscriptaddress,showpacs]{revtex4}
\usepackage[cp1250]{inputenc}
\usepackage{graphics}
\usepackage{color}

\usepackage[tbtags]{amsmath}

\usepackage[
pdftitle={Quantum tunneling of a single semifluxon in a 0-pi LJJ},
pdfauthor={Dr. E. Goldobin},
pdfsubject={..},
urlcolor=blue,
bookmarks=true,
bookmarksopen=true,
bookmarksnumbered=true
]{hyperref}

\include{MyMc}

\begin{document}

\title{%
  Quantum tunneling of a single semifluxon
  in a 0-$\pi$ Josephson junction
}

\author{E.~Goldobin}
\email{gold@uni-tuebingen.de}
\affiliation{
  Physikalisches Institut,
  Center for Collective Quantum Phenomena,
  Universit\"at T\"ubingen,
  Auf der Morgenstelle 14,
  D-72076 T\"ubingen, Germany
}

\author{K. Vogel}
\author{W. P. Schleich}
\affiliation{
 Universit\"at Ulm,
 Institut f\"ur Quantenphysik,
 D-89069 Ulm, Germany
}

\author{D.~Koelle}
\author{R.~Kleiner}
\affiliation{
  Physikalisches Institut -- Experimentalphysik II,
  Universit\"at T\"ubingen,
  Auf der Morgenstelle 14,
  D-72076 T\"ubingen, Germany
}

\pacs{
  74.50.+r,   
  75.45.+j,   
  85.25.Cp    
  03.65.-w    
}

\keywords{
  Long Josephson junction, sine-Gordon, fractional Josephson vortex,
  macroscopic quantum effects
}

\begin{abstract}

  We consider a symmetric 0-$\pi$ Josephson junction of length $L$, which classically can be in one of two degenerate ground states \state{s} or \state{a}, corresponding to supercurrents circulating clockwise or counterclockwise around the 0-$\pi$ boundary. When the length $L$ of the junction becomes smaller than the Josephson penetration depth $\lambda_J$, the system can switch from one state to the other due to thermal fluctuations or quantum tunneling. We map this problem to the dynamics of a single particle in a periodic double well potential and estimate parameters for which macroscopic quantum coherence may be observed. We conclude that this system is not very promising to build a qubit because (a) it requires very low temperatures to reach the quantum regime, (b) its tiny flux is hard to read out and (c) it is very sensitive to the asymmetries between the 0 and $\pi$ parts of the junction.

\end{abstract}

\date{\today}

\maketitle

\section{Introduction}
\label{Sec:Intro}

Josephson junctions with the phase drop of $\pi$ in the ground state ($\pi$ JJs)\cite{Bulaevskii:pi-loop} are intensively investigated as they promise important advantages for Josephson junction based electronics\cite{Terzioglu:1997:CompJosLogic,Terzioglu:1998:CJJ-Logic}, and, in particular, for JJ based qubits\cite{Ioffe:1999:sds-waveQubit,Yamashita:2005:pi-qubit:SFS+SIS,Yamashita:2006:pi-qubit:3JJ}. Nowadays a variety of technologies allow to manufacture such junctions\cite{Ryazanov:2001:SFS-PiJJ,Kontos:2002:SIFS-PiJJ,Tsuei:Review,Lombardi:2002:dWaveGB,Weides:2006:SIFS-HiJcPiJJ}.

One can also fabricate so-called 0-$\pi$ long Josephson junctions (0-$\pi$ LJJs)\cite{Bulaevskii:0-pi-LJJ,Smilde:ZigzagPRL,Hilgenkamp:zigzag:SF,Weides:2006:SIFS-0-pi}, i.e., LJJs some parts of which behave as 0 junctions and other parts as $\pi$ junctions. The most interesting fact about such junctions is that a semifluxon\cite{Goldobin:SF-Shape,Xu:SF-Shape}, i.e. a vortex of supercurrent, carrying \emph{one half} of the magnetic flux quantum $\Phi_0\approx2.07\times10^{-15}\,{\rm Wb}$, can be formed at the boundaries between the 0 and $\pi$ regions provided the JJ is long enough. Classically, such a 0-$\pi$ LJJ has a degenerate ground state corresponding to either positive or negative polarity of the semifluxon, which we denote as the \state{s} or \state{a} state, respectively. For these two polarities the circulations of the supercurrents and, therefore, the resulting magnetic fields, have different directions. The classical properties of semifluxons are under intense theoretical and experimental investigations\cite{Kogan:3CrystalVortices,Kirtley:SF:HTSGB,Kirtley:SF:T-dep,Hilgenkamp:zigzag:SF,Kirtley:IcH-PiLJJ,Goldobin:SF-ReArrange,Stefanakis:ZFS/2,Zenchuk:2003:AnalXover,Goldobin:Art-0-pi,Susanto:SF-gamma_c,Goldobin:2KappaGroundStates,Goldobin:F-SF,Kirtley:2005:AFM-SF,Weides:2006:SIFS-0-pi,Buckenmaier:2007:ExpEigenFreq,Nappi:2007:0-pi:Fiske}. While the classical properties of semifluxons (at least for systems with few semifluxons) are more or less understood, their quantum behavior and their possible applications in the quantum domain still have to be studied.

When the energy barrier separating the two degenerate classical ground states is small enough, the system may switch from one state to the other due to thermal excitation over the barrier or due to quantum tunneling through the barrier. Thermally induced flipping of a single semifluxon was already observed\cite{Kirtley:SF:T-dep}. Macroscopic quantum tunneling (MQT) and macroscopic quantum coherence (MQC) in the system of two coupled semifluxons in a 0-$\pi$-0 LJJ was investigated theoretically \cite{Kato:1997:QuTunnel0pi0JJ,Goldobin:2005:MQC-2SFs}. In Ref.~\onlinecite{Goldobin:2005:MQC-2SFs} it was also concluded that, unlike a \emph{pair} of semifluxons, a \emph{single} semifluxon in a 0-$\pi$ LJJ with moderate or large length is always in the classical regime.

In the present paper, we study a 0-$\pi$ JJ of length $L\lesssim\lambda_J$, which admits semifluxon-like solutions for any junction length provided the lengths and critical current densities of the 0 and $\pi$ parts are equal. We map the full problem to the dynamics of a single particle in a double well potential with periodic boundary conditions and estimate the crossover temperature as well as the tunneling rate (energy level splitting) between the states \state{s} and \state{a}.

\section{Model}
\label{Sec:Model}

\begin{figure}[!b]
  \includegraphics{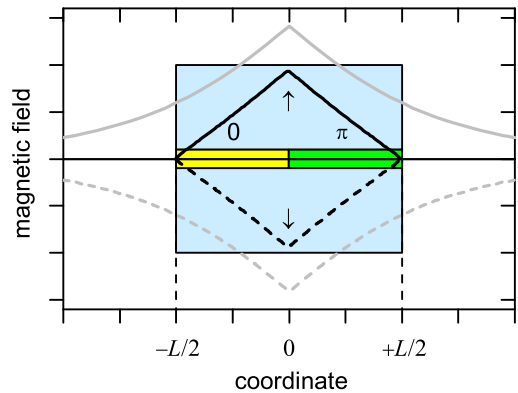}
  \caption{(Color online)
    Schematic drawing of a 0-$\pi$ JJ and magnetic field profiles in two degenerate ground states. Gray curves show the magnetic field profile in an infinite JJ (solid line is semifluxon, dashed line is antisemifluxon, $\Phi=\pm\Phi_0/2$). Black curves show corresponding solutions in the JJ of finite length. In this case the magnetic flux $|\Phi|<\Phi_0/2$.
  }
  \label{Fig:sketch}
\end{figure}

We consider a ``long'' one dimensional Josephson junction of length $L$, see Fig.~\ref{Fig:sketch}. The Josephson phase $\mu(x,t)$ is a continuous function of the position $x$  ($-L/2<x<+L/2$) measured along the JJ and of time $t$. We restrict ourselves to an undriven dissipationless system. The dynamics of such a system is described by the Lagrangian ${\cal L}=K-U$, where
\begin{equation}
  K = E_J \int_{-L/2}^{+L/2}
    \omega_p^{-2}\frac{\mu_t^2}{2}
  \,dx
  \label{Eq:K}
\end{equation}
represents the kinetic energy and

\begin{equation}
  U = E_J \int_{-L/2}^{+L/2}  \left\{
    \lambda_J^2\frac{\mu_x^2}{2}
    + 1-\cos[\mu(x,t)+\theta(x)] 
  \right\}\,dx
  , \label{Eq:U}
\end{equation}
is the potential energy. The subscripts $x$ and $t$ denote the partial derivatives with respect to position and time, respectively. In the above equations the three physical parameters are the Josephson energy per unit of junction length $E_J$, the Josephson penetration depth $\lambda_J$ and the Josephson plasma frequency $\omega_p$. The function $\theta(x)$ describes the position of 0  and $\pi$ regions along the junction. It is zero along 0 regions and is equal to $\pi$ along $\pi$ regions.

Applying the Euler-Lagrange formalism to our Lagrangian, we find that on the classical level the dynamics of the Josephson phase is described by the time-dependent sine-Gordon equation\cite{Goldobin:SF-Shape}
\begin{equation}
  \lambda_J^2\mu_{xx} - \omega_p^{-2} \mu_{tt} - \sin[\mu(x,t)+\theta(x)] = 0
  . \label{Eq:sG:time}
\end{equation}

\section{A fractional vortex in a symmetric 0-$\pi$ junction}
\label{Sec:0-pi-JJ}

Let us consider a symmetric 0-$\pi$ LJJ of the length $L$ ($-L/2<x<+L/2$) with a 0-$\pi$ boundary at $x=0$, see Fig.~\ref{Fig:sketch}. In this case $\theta(x)$ is a step function
\begin{equation}
  \theta(x) =
    \left\{
      \begin{array}{lrcl}
       0,   \quad & -L/2<&x&<0;\\
       \pi, \quad &    0<&x&<+L/2.
      \end{array}
    \right.
   \label{Eq:theta1}
\end{equation}

Classically, for any $L$ the trivial stationary solutions $\mu(x)=0$ or $\mu(x)=\pi$ are unstable and correspond to the Josephson energy maximum $U_\mathrm{max}=E_J L$, see Eq.~(\ref{Eq:U}). The stable stationary solutions depend on the length $L$ of the system.

If $L\gg\lambda_J$, the ground state of this system is a single semifluxon.\cite{Bulaevskii:0-pi-LJJ,Xu:SF-Shape,Goldobin:SF-Shape,Goldobin:SF-ReArrange} Such a semifluxon may have positive or negative polarity that corresponds to two classical degenerated states \state{s} and \state{a}\cite{Goldobin:SF-Shape,Goldobin:SF-ReArrange}, see the gray curves in Fig.~\ref{Fig:sketch}. In this limit the energy barrier separating two classical states is very large and the system is always in the classical regime\cite{Goldobin:2005:MQC-2SFs}. Therefore, in the present paper we investigate 0-$\pi$ LJJs with $L\lesssim\lambda_J$.

For $L\lesssim\lambda_J$ the stationary stable solutions of Eq.~(\ref{Eq:sG:time}) can be approximated by\cite{Buzdin:2003:phi-LJJ,Buzdin:2003:phi-LJJ:Bug} 
\begin{equation}
  \mu_0^\pm(x) \approx
  \pm\frac{\pi}{2} \pm \frac{x(L-|x|)}{2\lambda_J^2}
  \label{Eq:mu0}
\end{equation}
and correspond to the energy minima. Magnetic field profiles are shown in Fig.~\ref{Fig:sketch} by black lines. Substituting the stable and the unstable stationary solutions for $\mu$ into Eq.~(\ref{Eq:U}), we find that the barrier between two stationary stable solutions (\ref{Eq:mu0}) is given by
\begin{equation}
  U_0(\ell) \approx E_J \lambda_J \frac{\ell^3}{24}
  ,\label{Eq:DeltaU}
\end{equation}
where $\ell=L/\lambda_J$ is the normalized JJ length. In each of these ground states the flux in the junction is 
\begin{equation}
  \Phi = \frac{\Phi_0}{2\pi}\Delta\mu
  =\frac{\Phi_0}{2\pi}\left[\mu(L/2)-\mu(-L/2)\right]
  =\pm\Phi_0\frac{\ell^2}{8\pi}
  . \label{Eq:Phi}
\end{equation}

The main assumption of this paper is that for sufficiently short LJJ and sufficiently low energies the time dependent phase $\mu(x,t)$ can be approximated by (see Appendix \ref{App:CollCoord})
\begin{equation}
  \mu(x,t) =
  Q(t) + \frac12 \sin[Q(t)]\frac{x(L-|x|)}{\lambda_J^2}
  , \label{Eq:mu-Ansatz}
\end{equation}
i.e. it moves along a \emph{collective coordinate} $Q$. Note that the ansatz (\ref{Eq:mu-Ansatz}) satisfies zero magnetic field boundary conditions $\mu_x(\pm L/2)=0$. The values of $Q=\pm\pi/2$ correspond to the two distinct classical ground states (\ref{Eq:mu0}), whereas the values $Q=0$ or $Q=\pi$ correspond to unstable solutions $\mu=0$ and $\mu=\pi$.

By substituting the ansatz (\ref{Eq:mu-Ansatz}) into Eq.~(\ref{Eq:U}) we calculate the potential energy of the system as a function of $Q$ and obtain
\begin{equation}
  U(Q) = E_J \lambda_J \ell \left( 1-\frac{\ell^2}{24}\sin^2 Q \right)
  . \label{Eq:U(Q)}
\end{equation}
Clearly, the barrier between the two classical ground states is given by Eq.~(\ref{Eq:DeltaU}) which simultaneously is the amplitude of the $\cos$-like potential.

In essence, our ansatz (\ref{Eq:mu-Ansatz}) maps the full problem to the motion of a fictitious particle in the one-dimensional potential $U(Q)$ (\ref{Eq:U(Q)}).

To find the mass of this particle we substitute the ansatz (\ref{Eq:mu-Ansatz}) into Eq.~(\ref{Eq:K}) and find
\begin{equation}
  K(\dot{Q}) \approx \frac{E_J \ell}{\omega_p^2}\frac{\dot{Q}^2}{2}
  . \label{Eq:K(P)}
\end{equation}
Since $K(\dot{Q})\equiv M \dot{Q}^2/2$, the mass of our fictitious particle is given by
\begin{equation}
  M = \frac{E_J\lambda_J}{\omega_p^2}\ell\quad[\mathrm{kg\cdot m^2}]
  . \label{Eq:M}
\end{equation}
Note, that actually $M$ is not a mass, but a moment of inertia, since our coordinate $Q$ is not a position, but a dimensionless phase.

\subsection{Estimate of quantum-to-classical crossover}
\label{Sec:Est}

For a harmonic oscillator with mass $M$ and frequency $\omega_0$ the (square of the) width of the probability distribution in the ground state is determined by
\begin{equation}
  \langle \Delta Q^2 \rangle  = \frac{\hbar}{M\omega_0}.
  \label{Eq:HO-width}
\end{equation}
Our potential $U(Q)$, Eq.~(\ref{Eq:U(Q)}), is not parabolic, but we may approximate each potential well with a minimum at $Q_0=\pm\pi/2$ by a parabolic potential, i.e.,
\begin{equation}
  U(Q)\approx E_J\lambda_J\left[ 
    \ell-\frac{\ell^3}{24}+\frac{\ell^3}{24}(Q-Q_0)^2 
  \right]
  , \label{Eq:U@min}
\end{equation}
which together with the mass $M$ given by Eq.~(\ref{Eq:M}) describes a harmonic oscillator with the ``spring constant'' $k=E_J\lambda_J\ell^3/12$ and the eigenfrequency 
\begin{equation}
  \omega_0 = \sqrt{\frac{k}{M}}=\omega_p\frac{\ell}{2\sqrt{3}}
  . \label{Eq:omega_0}
\end{equation}

Using our expressions for the mass $M$ (\ref{Eq:M}), the frequency $\omega_0$ (\ref{Eq:omega_0}), and the harmonic oscillator width (\ref{Eq:HO-width}) we obtain
\begin{equation}
  \langle {\Delta Q}^2 \rangle =
  \frac{\hbar\omega_p}{E_J\lambda_J}
  \frac{2\sqrt{3}}{\ell^2}
  \label{Eq:Q2av}
\end{equation}
as an estimate for the spread of a wave function in each of the wells of the potential $U(Q)$. Quantum effects are noticeable when the wave function in the left well overlaps with the wave function in the right well. This overlap should be appreciable, but not too large since otherwise the two states will not be distinguishable. For a rough estimate we take   $\langle {\Delta Q}^2 \rangle \gtrsim 0.1 Q_0^2$ as criterion for quantum behavior. Thus quantum effects will dominate if 
\begin{equation}
  \begin{split}
    \frac{\langle {\Delta Q}^2 \rangle}{Q_0^2}
    &=\frac{8\sqrt{3}}{\pi^2}\frac{\hbar\omega_p}{E_J\lambda_J}
    \frac{1}{\ell^2} \\
    &= \frac{32\sqrt{3}}{1}
    \frac{\hbar}{\Phi_0^2}\sqrt{\frac{\mu_0d'}{C w^2}}
    \frac{1}{\ell^2}\gtrsim 0.1
    , \label{Eq:QuDomination}
  \end{split}
\end{equation}
where we took into account the definitions
\begin{equation}
  \lambda_J = \sqrt{\frac{\Phi_0}{2\pi\mu_0 d' j_c}},\;
  \omega_p  = \sqrt{\frac{2\pi j_c}{\Phi_0 C}},\;
  E_J       = \frac{j_c w \Phi_0}{2\pi}
  . \label{Eq:def}
\end{equation}
In Eq.~(\ref{Eq:def}), $\mu_0d'$ is the inductance per square of the superconducting electrodes ($\mu_0$ is the permeability of vacuum, $d' \approx 2\lambda_L$, $\lambda_L$ is the London penetration depth), $j_c$ is the critical current density of the LJJ, $C$ is the capacitance of the LJJ per unit of area, and $w$ is the LJJ's width.

For typical parameters $\lambda_L=90\units{nm}$, $w=1\units{\mu m}$ and $C=4.1\units{\mu F/cm^2}$ (\textsc{Hypres}\cite{Hypres} technology with $j_c=100 \units{A/cm^2}$) Eq.~(\ref{Eq:QuDomination}) reads
\begin{equation}
  \frac{\langle {\Delta Q}^2 \rangle}{Q_0^2}
  \approx 3.2\times10^{-3}{\ell}^{-2}\gtrsim 0.1
  . \label{Eq:EstSpread}
\end{equation}
Thus, quantum effects start to play a role for ${\ell}\lesssim0.18$. Note, that according to Eq.~(\ref{Eq:QuDomination}) the classical-to-quantum crossover length does not depend on $j_c$.

Using definitions (\ref{Eq:def}) in terms of physical parameters of the LJJ, we can express the inertial mass $M$ (\ref{Eq:M}) as
\begin{equation}
  M \approx \ell
  \frac{wC}{\sqrt{\mu_0 d' j_c}}
  \left( \frac{\Phi_0}{2\pi} \right)^\frac52
  \approx 1.3\times10^{-4} \ell m_e \lambda_J^2
  . \label{Eq:M:ph}
\end{equation}
This means that a single electron moving around the whole JJ (circulating around 0-$\pi$ boundary) has much larger moment of inertia than our semifluxon.

To estimate the classical-to-quantum crossover temperature, we compare the thermal escape rate and the semiclassical expression for the quantum mechanical energy level splitting. The thermal escape rate is $\propto \exp(-U_0/k_BT)$, while in the semiclassical limit the energy level splitting is $\propto\exp(-4 U_0/\hbar\omega_0)$, see Eq.~(\ref{eq:delta-periodic}). Neglecting prefactors we conclude that at the temperature
\begin{equation}
  T_x \sim \frac{\hbar\omega_0}{4 k_B}
  = \frac{\hbar\omega_p\ell}{8\sqrt3 k_B}
  \label{Eq:Tx}
\end{equation}
quantum tunneling and thermal escape have the same rate, while at lower temperature quantum tunneling dominates. For the parameters chosen above and $\ell=0.18$ this gives $T_x\approx 27\units{mK}$.

Another specific temperature $T^\star$ is defined by comparing $k_BT$ with the barrier height. For $T>T^\star$ the states \state{s} and \state{a} cannot be distinguished, i.e., even in the classical regime read out cannot be properly organized. For our Josephson junction the temperature $T^\star$ is given by
\begin{equation}
  T^\star = \frac{U_0}{k_B} = \frac{E_J \lambda_J}{k_B}\frac{\ell^3}{24}
  . \label{Eq:T*}
\end{equation}
For $\ell=0.18$ we obtain $T^\star\approx 218\units{mK}$. Note that $T^\star$ can become smaller than $T_x$ for small values of $\ell$. For such parameters, however, we have $U_0 < \hbar\omega$ and we are not in the semiclassical limit used to derive Eq.~(\ref{Eq:Tx}).

\subsection{Energy level splitting}

The stationary Schrödinger equation for our collective coordinate $Q$ reads
\begin{equation}
  \left[ -\frac{\hbar^2}{2M}\fracp[2]{}{Q} + U(Q) \right]\psi = E \psi
  . \label{Eq:Schroedinger0}
\end{equation}
Since in our configuration the Josephson phases $\mu + 2\pi$ and $\mu$ cannot be distinguished, $\mu$ (and therefore also $Q$) is only defined modulo $2\pi$. Therefore, we supplement Eq.~(\ref{Eq:Schroedinger0}) with periodic boundary conditions $\psi(Q+2\pi) = \psi(Q)$.

The main purpose of our quantum mechanical calculations is to
investigate quantum tunneling between the two classical ground
states. More precisely, we want to approximate our Josephson junction
by a quantum mechanical two-level system which may be used to implement
a qubit. In such a two-level system the dynamics of the two lowest
levels is decoupled from the other levels. This requires that the
two lowest energy eigenvalues $E_1$ and $E_2$ of the Schrödinger
equation~(\ref{Eq:Schroedinger0}) are well-separated from the other energy eigenvalues.

Furthermore, we want to observe coherent oscillations between
the left and the right well of the potential.
This is only possible if there are superpositions of the ground state $\psi_1(Q)$ and the first excited state $\psi_2(Q)$ of the Schrödinger equation~(\ref{Eq:Schroedinger0}) which are sufficiently well localized in the left and in the right potential well.

We expect that we can approximate our system by a quantum
mechanical two-level system, if the two classical (degenerate) ground
states are separated from each other by a sufficiently high energy barrier. In our system, the barrier height depends on the scaled length $\ell$ of the Josephson junction, see Eq.~(\ref{Eq:DeltaU}). To gain more insight, we have solved our Schrödinger equation~(\ref{Eq:Schroedinger0}) numerically for different values of $\ell$. The parameters of the Josephson junction are given in Sec.~\ref{Sec:Est} above Eq.~(\ref{Eq:EstSpread}).

\begin{figure}[!htb]
  \includegraphics{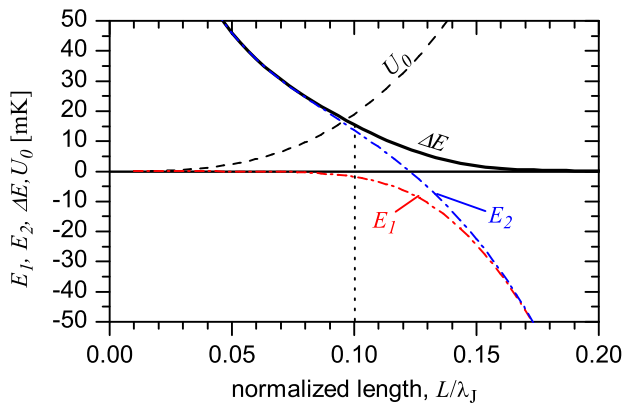}
  \caption{%
    The two lowest energy levels $E_1$ and $E_2$, $\Delta E=E_2-E_1$ and the barrier height $U_0$ (all given in mK) as a function of $\ell$ for typical parameters given in the text.
  }
  \label{Fig:dE(ell)}
\end{figure}

\begin{figure}[!htb]
  \includegraphics{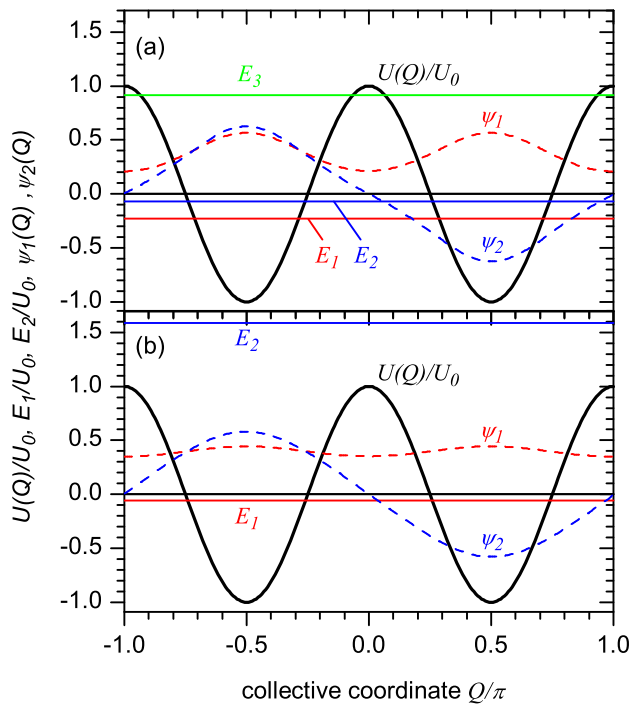}
  \caption{%
    Solutions of the Schrödinger equation~(\ref{Eq:Schroedinger0}) for (a) $\ell=0.13$, (b) $\ell=0.09$.
  }
  \label{Fig:Solutions}
\end{figure}

Fig.~\ref{Fig:dE(ell)} shows the lowest energy levels $E_1$ and $E_2$, the energy level splitting $\Delta E = E_2 -E_1$, and the barrier height $U_0$
of the potential as a function of the normalized JJ length $\ell$.
When we reduce $\ell$, the lowest energy eigenvalues split. For
$\ell \lesssim 0.1$ the energy level $E_2$ is lifted above $U_0$ so
that we have only one energy level below $U_0$.

Fig.~\ref{Fig:Solutions} shows the eigenfunctions $\psi_1(Q)$ and $\psi_2(Q)$ corresponding to the eigenvalues $E_1$ and $E_2$. We have chosen two values for $\ell$ which lead to a qualitatively different behavior.

For $\ell = 0.13$ (Fig.~\ref{Fig:Solutions}a) we have a
quantum mechanical two-level system as discussed above: The two
lowest energy eigenvalues $E_1$ and $E_2$ are well-separated from
the others and $\left[\psi_1(Q) \pm \psi_2(Q)\right]/\sqrt{2}$
is well-localized in one of the potential wells.

For $\ell=0.09$ (Fig.~\ref{Fig:Solutions}b) our simple picture of a quantum mechanical two-level system fails: The difference between $E_1$ and $E_2$ is large and only one energy eigenvalue is below $U_0$. Furthermore, the two lowest eigenvalues are not well-separated from the higher energy eigenvalues (not shown in Fig.~\ref{Fig:Solutions}b). The ground state $\psi_1(Q)$ is completely delocalized and $|\psi_1(Q)|^2$ does not have pronounced maxima at the minima of the potential. Obviously, for $\ell=0.09$ we cannot approximate our system by a two-level system.

As expected, only for sufficiently large values of $\ell$, i.e., sufficiently high energy barriers, our Josephson junction behaves like a two-level system.  However, we have to take into account that thermal excitations will destroy quantum coherence if the energy level splitting $\Delta E$ becomes too small. Present technology restricts $\Delta E$ to values larger than $20\ldots30\units{mK}$ --- the temperature achievable in modern dilution refrigerators. According to Fig.~\ref{Fig:dE(ell)} this requires $\ell < 0.09$. For such values of $\ell$, we are not in the limit of a two-level system, see Fig.~\ref{Fig:Solutions}b.

\section{Discussion}
\label{Sec:Disc}

\begin{table}
  (a) single semifluxon\\
  \begin{tabular}{r | r r r r r}
    $\ell$ & $T^{\star}$ & $T_x$
    & $\Delta E_{12}^\mathrm{sc}$ 
    & $\Delta E_{12}^\mathrm{num}$ & $\Delta E_{23}^\mathrm{num}$ \\
    \hline
    0.18 & 218 & 27.0 &   0.3 &  0.2 & 89.6\\
    0.13 &  83 & 19.5 &   7.2 &  5.3 & 40.8\\
    0.10 &  38 & 15.0 &  24.4 & 15.5 & 18.9\\
    0.09 &  28 & 13.5 &  32.0 & 19.9 & 18.8\\
    \hline
  \end{tabular}
  \\[5mm]
  (b) AFM semifluxon molecule\\
  \begin{tabular}{r | r r r r r}
    $\delta a$ & $T^{\star}$ & $T_x$ 
    & $\Delta E_{12}^\mathrm{sc}$ 
    & $\Delta E_{12}^\mathrm{num}$ & $\Delta E_{23}^\mathrm{num}$ \\
    \hline
    0.020 & 566 & 58 &  0.2 &  0.2 & 265\\
    0.015 & 318 & 51 &  3.4 &  2.8 & 183\\
    0.010 & 141 & 41 & 36.5 & 22.9 & 124\\
    \hline
  \end{tabular}
  \caption{%
    The crossover temperatures $T^{\star}$ and $T_x$, the energy level splitting $\Delta E_{12} = E_2-E_1$, and the energy difference $\Delta E_{23}=E_3-E_2$ between the first and the second excited state for a single semifluxon (a) and a semifluxon molecule (b) for different values of the scaled lengths $\ell$ and $\delta a$. All values are given in mK. The numbers for the semiclassical energy level splitting $\Delta E_{12}^\mathrm{sc}$ are based on
    Eq.~(\ref{eq:delta-periodic}) (a) and Eq.~(\ref{eq:delta-quartic}) (b), whereas $\Delta E_{12}^\mathrm{num}$ and $\Delta E_{23}^\mathrm{num}$ were calculated numerically. For $\Delta E_{12} \ll \Delta E_{23}$ the systems are good two-level systems.
  }
  \label{Tab:CrossOver}
\end{table}

The results for the parameters used in Fig.~\ref{Fig:dE(ell)} and Fig.~\ref{Fig:Solutions} are not very promising: either we are not in the limit of a two-level system or we need temperatures significantly lower than $20\ldots30\units{mK}$ to avoid thermal excitations. We can try to improve the situation by using different parameters for the Josephson junction.

In Appendix~\ref{App:Mathieu} we show that the Schrödinger equation~(\ref{Eq:Schroedinger0}) is equivalent to the Mathieu equation~(\ref{eq:mathieu}) with only one scaled system parameter $h$ and the scaled energy $\varepsilon$, see Eqs.~(\ref{eq:h-def}) and (\ref{eq:epsilon-def}). Using the definitions~(\ref{Eq:def}), we find that the parameter $h$ is proportional to $(j_c w)^2$ and that the scaling factor between the energy $E$ and the scaled energy $\varepsilon$ is proportional to $w^{-1}$ and does not depend on $j_c$. If we reduce the width $w$ by a factor of $10$ and simultaneously increase $j_c$ by a factor of $10$, we can increase the energy scale by a factor of $10$ without changing the generic behavior of the system ($h$ does not change). For the parameters used in Fig.~\ref{Fig:Solutions}a we could increase the energy level splitting from $5.3\units{mK}$ to $53\units{mK}$ and would still be in the limit of a two-level system.

To read out the state of a semifluxon one can use a SQUID situated
just in front of the 0-$\pi$ boundary. Note, however, that the two states
carry the flux $\Phi\ll \Phi_0/2$. For small $\ell$ the flux is given by Eq.~(\ref{Eq:Phi}) and is equal to $\sim 10^{-3}\Phi_0$ for
$\ell=0.18$. Needless to say that it is rather difficult to measure such a small flux accurately.

Another problem is a high sensitivity to the parameter spread. All calculations done so far are for a symmetric 0-$\pi$ junction, i.e., for the ideal case where the critical current densities and lengths of the 0 and $\pi$ parts are equal. In experiments, though, a small asymmetry is always present. For $\ell \ll 1$ this may easily lead to the situation where the classical ground state is just $\mu=0$ (when the 0 part is longer or has a higher $j_c$) or $\mu=\pi$ (when the $\pi$ part is longer or has a higher $j_c$) which is not doubly degenerate. For $\ell \ll 1$ the length asymmetry should not exceed\cite{Buzdin:2003:phi-LJJ} $\ell^3/48$, which is $\sim 1.2 \times 10^{-4}$ for $\ell=0.18$. Such precise control of LJJ length is not feasible.


Finally we want to compare our results to the results we obtained for an AFM semifluxon molecule in a $0$-$\pi$-$0$ junction \cite{Goldobin:2005:MQC-2SFs}. In such a system the main parameter which controls the height of the energy barrier (similar to $\ell$ in this paper) is the length $a$ of the $\pi$ region. It turns out that to approach the quantum regime $a$ should be just a bit above a crossover length $a_c=\pi\lambda_J/2$, i.e., $a=(\pi/2+\delta a)\lambda_J$. The corresponding key numbers are given in Tab.~\ref{Tab:CrossOver}b. According to the Tab.~\ref{Tab:CrossOver} the semifluxon molecule in a $0$-$\pi$-$0$ junction is more suitable to build qubits. The crossover temperatures for $\delta a = 0.01$ can be reached with modern dilution refrigerators.

\section{Conclusions}
\label{Sec:Concl}

We have presented a simple quantum theory of a short 0-$\pi$ JJ. By introducing a collective coordinate, the full problem was reduced to a dynamics of a single particle in a one-dimensional potential.

It was found that the quantum regime is reached when the normalized length $\ell$ of the junction is smaller than 0.18. The classical-to-quantum crossover temperature $T_x$ is about $20\units{mK}$ for the case when we still have a good two-level system and the semiclassical approximation (used to define $T_x$) is still valid. This is on the limit of modern dilutions refrigerators. Moreover, even for $\ell\leq0.18$ the flux, which should be read out to determine the state of the system, is $\lesssim 10^{-3}\Phi_0$, which is rather difficult to detect even using SQUIDs. The system is also very sensitive to  asymmetries in the 0 and $\pi$ parts, which should not exceed $10^{-4}$.

It is quite interesting that another system, the so-called "d-dot"\cite{Koyama:2005:d-dot:Qu}, is equivalent to the 0-$\pi$ LJJ of the present paper. A "d-dot" is the squared island of a $d$-wave superconductor embedded into an $s$-wave subspace. Essentially it is an annular 0-$\pi$-0-$\pi$ LJJ with the length of each region equal to $a$. The two ground states are \state{sasa} and \state{asas}. In fact, the four regions of a $d$-dot were used just because of the topological limitations. One can use symmetry arguments to conclude that such an annular 0-$\pi$-0-$\pi$ JJ is equivalent in terms of ground states and eigenmodes to an annular 0-$\pi$ JJ with only two facets of length $a$, which, in turn, is equivalent to a linear (open ended) 0-$\pi$ JJ with the total length $a$. In contrast to Ref.~\onlinecite{Koyama:2005:d-dot:Qu}, we conclude that such a system is not a very promising candidate to build qubits. Better figures are shown by a two-semifluxon molecule in a 0-$\pi$-0 JJ  with tunneling between \state{sa} and \state{as} states\cite{Kato:1997:QuTunnel0pi0JJ,Goldobin:2005:MQC-2SFs}.

\begin{acknowledgments}

We thank E.~Il'ichev and F.~Wilhelm for fruitful discussions. Financial support by DFG (project SFB/TRR-21) is greatfully acknowledged.
  
\end{acknowledgments}

\appendix

\section{Derivation of collective coordinate}
\label{App:CollCoord}

In this appendix we motivate the use of the collective coordinate
$Q(t)$, that is, the assumption that for sufficiently short LJJs
and sufficiently low energies we can approximate the Josephson
phase $\mu(x,t)$ by
\begin{equation}
\mu(x,t) \approx Q(t) + \sin [Q(t)] \frac{x \left(L-|x|\right)}{2\lambda_J^2} ,
\end{equation}
where $Q(t)$ describes the motion of a single particle of mass
\begin{equation}
M = \frac{E_J \lambda_J}{\omega_p^2} \, \ell
\end{equation}
in the potential
\begin{equation}
U(Q) = E_J \lambda_J \ell
       \left( 1 - \frac{\ell^2}{24} \sin^2 Q \right) .
\end{equation}

In order to keep the notation simple, we use the scaled length
$\ell = L/\lambda_J$ of the Josephson junction, the scaled time
$\tau = \omega_p t$ and the scaled position $\xi = x/\lambda_J$
in the rest of this appendix.

\subsection{Fourier expansion}
\label{Sec:Fourier}

We start from the Lagrangian
\begin{eqnarray}
\label{eq:L-field}
{\cal L} = E_J \lambda_J \int\limits_{-\ell/2}^{+\ell/2} \bigg\{
        \frac{1}{2} \left(\frac{\partial \mu}{\partial \tau}\right)^2
      - \frac{1}{2} \left(\frac{\partial \mu}{\partial \xi}\right)^2
\nonumber\\
      - \left[ 1 - \cos\left(\mu + \theta\right) \right]
    \bigg\} \, d\xi
\end{eqnarray}
for the Josephson phase $\mu(\xi,\tau)$ and expand $\mu(\xi,\tau)$
into a Fourier series which takes into account the boundary conditions
$\mu_{\xi}(-\ell/2,\tau) = \mu_{\xi}(\ell/2,\tau) = 0$. The most general
Fourier series that takes into account these boundary conditions is
\begin{eqnarray}
&& \mu(\xi,\tau) = \,Q(\tau)
+ \sum\limits_{n=1}^{\infty}
q^{(c)}_n(\tau) \cos\left(\frac{2n}{\ell}\,\pi\xi\right)
\nonumber\\
&& \hspace*{4em} + \sum\limits_{n=0}^{\infty} q_n^{(s)}(\tau)
\sin\left(\frac{2n+1}{\ell}\,\pi\xi\right) .
\end{eqnarray}

In the following calculations we make various approximations for small
$\ell$. A detailed analysis shows that the cosine terms in this Fourier
series only lead to higher-order corrections in $\ell^2$ which are to
a large extent not consistent with the approximations we are going to make.

In order to keep the calculations brief, we therefore substitute
the Fourier series
\begin{equation}
\label{eq:fourier1}
\mu(\xi,\tau) = \,Q(\tau)
+ \sqrt{2}\sum\limits_{n=0}^{\infty} q_n(\tau)
\sin\left(\frac{2n+1}{\ell}\,\pi\xi\right)
\end{equation}
into the Lagrangian~(\ref{eq:L-field}). The factor $\sqrt{2}$
in front of the sum was chosen for convenience, see the kinetic
energy in Eq.~(\ref{eq:L-approx}).  For short LJJs, non-uniform excitations
require large energies since for small values of $\ell$ and significantly
large non-uniform excitations the term $\partial\mu/\partial\xi$ in
the potential energy of the Lagrangian~(\ref{eq:L-field}) becomes large.
Therefore, we expect that the Fourier coefficients $q_n(\tau)$ are small
for low energies. Since we are interested in the low-energy behavior of
the system, we only take into account terms in the Lagrangian that are
quadratic in $q_n(\tau)$ and and neglect higher-oder terms. For the
uniform part $Q(\tau)$, however, we make no assumption. A similar
approach was used by Koyama et al.~\cite{Koyama:2005:d-dot:Qu} for a square-shaped closed $0$-$\pi$ junction (d-dot).

Due to the orthogonality of the sine functions in the Fourier
series~(\ref{eq:fourier1}) the only non-trivial integral we have
to calculate is
\begin{equation}
\int\limits_{-\ell/2}^{+\ell/2}
\cos\left[\mu(\xi,\tau) + \theta(\xi)\right] \, d\xi .
\end{equation}
Here we make use of our assumption that the non-uniform part
$\delta(\xi,\tau) = \mu(\xi,\tau) - Q(\tau)$ of the Josephson phase
is small and rewrite $\cos\left[ \mu(\xi,\tau) + \theta(\xi) \right]$
in the form
\begin{eqnarray}
\label{eq:cos-approx}
\cos\left[ \mu(\xi,\tau) + \theta(\xi) \right]
& = & \cos Q(\tau) \cos\delta(\xi,\tau) \cos\theta(\xi)
\nonumber\\
&& - \sin Q(\tau) \sin\delta(\xi,\tau) \cos\theta(\xi)
\nonumber\\
& \approx & \cos Q(\tau) \, \left[1-\delta^2(\xi,\tau)/2\right]
            \cos\theta(\xi)
\nonumber\\
&& - \sin Q(\tau) \,\, \delta(\xi,\tau) \, \cos\theta(\xi) ,
\end{eqnarray}
where we have used $\sin\theta(\xi)=0$.

The integral over $\cos\theta(\xi)$ alone is zero since the
$0$-region and $\pi$ region have the same length.
We are therefore left with the two integrals
\begin{equation}
\int\limits_{-\ell/2}^{\ell/2} \delta(\xi,\tau) \cos\theta(\xi) \, d\xi
\end{equation}
and
\begin{equation}
\int\limits_{-\ell/2}^{\ell/2}  \delta^2(\xi,\tau) \cos\theta(\xi) \, d\xi .
\end{equation}
Due to the different sign of $\cos\theta(\xi)$ in the $0$- and in the
$\pi$ region we cannot use the orthogonality relations of the sine functions
in Eq.~(\ref{eq:fourier1}). Evaluating the integrals is nevertheless
straightforward as the second integral vanishes for symmetry reasons.

Within our approximation we finally arrive at the approximate Lagrangian
\begin{equation}
\label{eq:L-approx}
{\cal L} \approx E_J \lambda_J \ell \Big[
    \frac{1}{2}\dot{Q}^2
    + \frac{1}{2}\sum\limits_{n=0}^{\infty} \dot{q}_n^2
- V\left(Q,\left\{q_n\right\}\right)
\Big] ,
\end{equation}
where the dot denotes the derivative with respect to $\tau$ and where the
potential energy $V\left(Q,\left\{q_n\right\}\right)$ is given by
\begin{eqnarray}
\label{eq:V1}
&& V\left(Q,\left\{q_n\right\}\right)
= 1 +
\frac{\pi^2}{2\ell^2}\sum\limits_{n=0}^{\infty} (2n+1)^2 q_n^2
\nonumber\\
&& \hspace*{6.5em}
- \frac{2\sqrt{2}}{\pi} \sin Q \sum\limits_{n=0}^{\infty}
  \frac{q_n}{2n+1} .
\end{eqnarray}

\subsection{Harmonic oscillators for ${q_n}$}
\label{sec:HO}

For fixed values of $Q$ the potential
$V\left(Q,\left\{q_n\right\}\right)$ is harmonic with respect to
$\left\{q_n\right\}$. In order to get more insight, we rewrite it
in the form
\begin{equation}
\label{eq:V2}
V\left(Q,\left\{q_n\right\}\right) = V_0(Q)
+ \frac{\pi^2}{2\ell^2}\sum\limits_{n=0}^{\infty} (2n+1)^2
       \left[q_n-\bar{q}_n(Q)\right]^2 .
\end{equation}

By comparing Eqs.~(\ref{eq:V1}) and (\ref{eq:V2}) we find
that we have to choose
\begin{equation}
\label{eq:q-bar}
\bar{q}_n(Q) = \frac{2\sqrt{2}\,\ell^2}{\pi^3 (2n+1)^3} \sin Q
\end{equation}
and $V_0(Q)$ is given by
\begin{eqnarray}
\label{eq:Veff}
V_0(Q) &=& 1 - \frac{\pi^2}{2\ell^2}\sum\limits_{n=0}^{\infty}
           (2n+1)^2 \bar{q}_n^2(Q)
\nonumber\\
& = & 1 -\frac{\ell^2}{24} \sin^2 Q ,
\end{eqnarray}
where in the last step we have used~\cite{Prudnikov-1986}
\begin{equation}
\sum\limits_{n=0}^{\infty} \frac{1}{(2n+1)^4} = \frac{\pi^4}{96} .
\end{equation}

For fixed values of $Q$ this potential describes a sum of independent
harmonic oscillators with (scaled) frequencies
\begin{equation}
\Omega_n = \frac{(2n+1)\pi}{\ell} , \quad n = 0, 1, 2, 3 \dots .
\end{equation}

\subsection{Stationary solution}

The stationary solutions of the full potential
$V(Q,\{q_n\})$ follow from
\begin{subequations}
\begin{eqnarray}
\frac{\partial V}{\partial Q} & = &
\frac{dV_0}{dQ} - \frac{\pi^2}{\ell^2}\sum\limits_{n=0}^{\infty}
(2n\!+\!1)^2 \left[q_n\!-\!\bar{q}_n(Q)\right] \frac{d\bar{q}_n}{dQ} = 0 ,
\nonumber\\
\\
\frac{\partial V}{\partial q_n} &=&
\frac{\pi^2}{\ell^2} (2n\!+\!1)^2
\left[q_n-\bar{q}_n(Q)\right] = 0 .
\end{eqnarray}
\end{subequations}
From these conditions we immediately find
\begin{equation}
\sin Q_0 \cos Q_0 = 0 , \qquad q_n^0 = \bar{q}_n(Q_0) .
\end{equation}
We therefore have two classes of stationary solutions:
\begin{eqnarray}
\label{eq:stat1}
&& \sin Q_0 = 0, \quad Q_0 = k\pi
\quad (k=0,\pm 1, \pm 2, \dots) ,
\nonumber\\[2ex]
&& q_n^0  = \bar{q}_n(Q_0) = 0
\end{eqnarray}
and
\begin{eqnarray}
\label{eq:stat2}
&& \cos Q_0 = 0, \quad Q_0 = \pi/2  + k\pi
\quad (k=0,\pm 1, \pm 2, \dots) ,
\nonumber\\[2ex]
&& q_n^0  = \bar{q}_n(Q_0) =
\frac{(-1)^k2\sqrt{2}\,\ell^2}{\pi^3 (2n+1)^3} .
\end{eqnarray}

A stability analysis shows that the stationary solutions corresponding
to Eq.~(\ref{eq:stat1}) are unstable whereas the the stationary solutions
corresponding to Eq.~(\ref{eq:stat2}) are stable.

The first class of stationary solutions, Eq.~(\ref{eq:stat1}),
obviously corresponds to the uniform stationary Josephson phases
$\mu_k^0(\xi)=k\pi$. When we substitute Eq.~(\ref{eq:stat2}) into the
into the Fourier series~(\ref{eq:fourier1}) we obtain the non-uniform
stationary Josephson phases
\begin{eqnarray}
&& \mu_k^0(\xi) = \frac{\pi}{2} + k\pi
+ \frac{(-1)^k 4\ell^2}{\pi^3}
\nonumber\\
&& \hspace*{3.5em} \times
\sum\limits_{n=0}^{\infty} \frac{1}{(2n+1)^3}
\sin\left(\frac{2n+1}{\ell} \pi \xi\right)
\nonumber\\[2ex]
&& \hspace*{2.5em} =
\frac{\pi}{2} + k\pi + \frac{(-1)^k}{2}
\, \xi \left( \ell-|\xi| \right) ,
\end{eqnarray}
where we have used~\cite{Prudnikov-1986}
\begin{equation}
\label{eq:sum-sin3}
\sum\limits_{n=0}^{\infty}
\frac{\sin\left[(2n+1)x\right]}{(2n+1)^3}
= \frac{\pi}{8}\,x(\pi-|x|) , \quad
-\pi \le x \le \pi .
\end{equation}

\subsection{Reduction to a one-dimensional system}

\subsubsection{Harmonic oscillators at low energies}

In Sec.~\ref{sec:HO} we found that for fixed values of $Q$ the potential
$V(Q,\{q_n\})$ describes a sum of independent harmonic oscillators with
frequencies $\Omega_n = (2n+1)\pi/\ell, n = 0, 1, 2, 3, \dots$. For decreasing
$\ell$ the harmonic potential becomes steeper and steeper whereas the
barrier height in the potential $V_0(Q)$ decreases, see Eq.~(\ref{eq:Veff}).
Therefore, we expect that for small $\ell$ and low energies the dynamics of the full system is
restricted to small fluctuations around $\bar{q}_n(Q)$ and we may approximate
$q_n(\tau)$ by
\begin{equation}
q_n(\tau) \approx \bar{q}_n(Q(\tau)) \,.
\end{equation}

When we substitute this approximation into the Lagrangian~(\ref{eq:L-approx})
we obtain
\begin{equation}
{\cal L} \approx E_J \lambda_J \ell \bigg\{
    \frac{1}{2} \dot{Q}^2 \Big[
       1 + \sum\limits_{n=0}^{\infty} \left(\frac{d\bar{q}_n}{dQ}\right)^2
    \Big] - V_0(Q) \bigg\} .
\end{equation}
The terms with the derivatives of $\bar{q}_n(Q)$ are of the order
$\ell^4$ whereas in $V_0(Q)$ only the terms $\sim \ell^2$ are taken into account. Therefore, we decided to neglect
these terms and finally arrive at the effective Lagrangian
\begin{equation}
{\cal L}_{\textrm{eff}} \approx E_J \lambda_J \ell
   \left[
      \frac{1}{2} \dot{Q}^2 - V_0(Q)
   \right]
\end{equation}
which describes the motion of a particle of mass
\begin{equation}
\label{eq:mass}
M = \frac{E_J \lambda_J}{\omega_p^2} \, \ell
\end{equation}
in the potential
\begin{equation}
\label{eq:Ueff}
U(Q) = E_J \lambda_J \ell
      \left( 1 - \frac{\ell^2}{24} \sin^2 Q \right) .
\end{equation}
The dynamics of a particle in this potential can be characterized
by the barrier height
\begin{equation}
\label{eq:U_0}
U_0 = \frac{\ell^3}{24} E_J \lambda_J
\end{equation}
of the potential $U(Q)$ and the frequency
\begin{equation}
\label{eq:omega_0}
\omega_0 = \frac{\omega_p \ell}{2\sqrt{3}}
\end{equation}
for small oscillations around the minima of the potential $U(Q)$.

The Fourier series~(\ref{eq:fourier1}) connects $Q(\tau)$ and $q_n(\tau)$
with the Josephson phase $\mu(\xi,\tau)$. When we substitute
Eq.~(\ref{eq:q-bar}) into the Fourier series~(\ref{eq:fourier1})
the corresponding approximation for the Josephson phase $\mu(\xi,\tau)$
reads
\begin{eqnarray}
\label{eq:mu-Q}
\mu(\xi,\tau) &\approx& Q(\tau) + \frac{4\ell^2}{\pi^3} \sin Q(\tau)
\nonumber\\
&& \hspace*{3.5em} \times
\sum\limits_{n=0}^{\infty} \frac{1}{(2n+1)^3}
\sin\left(\frac{2n+1}{\ell} \,\pi \xi\right)
\nonumber\\[2ex]
&=&  Q(\tau) + \frac{1}{2} \sin Q(\tau)
\, \cdot \, \xi \left( \ell-|\xi| \right) ,
\end{eqnarray}
where in the last step we have used Eq.~(\ref{eq:sum-sin3}) to evaluate
the sum. Equation~(\ref{eq:mu-Q}) is the central result of this appendix
and is used as the starting point in Sec.~\ref{Sec:0-pi-JJ}, see Eq.~(\ref{Eq:mu-Ansatz}).

Finally we want to mention that our results are in agreement with
the assumption of Sec.~\ref{Sec:Fourier} that the non-uniform part
of the Josephson phase is small.

\subsubsection{Quantum mechanical considerations}

So far our discussions were purely classical. In the main part of the
paper, however, we use $Q(\tau)$ as a collective coordinate to
investigate quantum tunneling in Josephson junctions. We now briefly
estimate under which conditions our collective coordinate $Q(\tau)$
is useful in the quantum limit of the system.

We have several energy scales in the system: the thermal energy
$k_B T$, the barrier height $U_0$, Eq.~(\ref{eq:U_0}), of the
potential $U(Q)$, and the excitations $\hbar\omega_p\Omega_n$
of the harmonic oscillators discussed in the Sec.~\ref{sec:HO}. Using
$Q(\tau)$ as a collective coordinate in the quantum regime obviously only
makes sense if these harmonic oscillators are not excited. Therefore
$k_B T$ and $U_0$ have to be much smaller than $\hbar\omega_p\Omega_n$.
Since we are mainly interested in quantum tunneling in the potential
$U(Q)$, $k_BT$ has to be smaller than $U_0$, see also the discussion
in Sec.~\ref{Sec:Est}. The condition that the harmonic oscillators are not excited therefore reduces to
\begin{equation}
U_0 \ll \hbar\omega_p \Omega_0 = \hbar\omega_p \frac{\pi}{\ell}
\end{equation}
or
\begin{equation}
\ell^4 \ll \frac{24\pi\hbar\omega_p}{E_J \lambda_J} .
\end{equation}
Furthermore, we have to restrict ourselves to $\ell^2 \ll 1$ since
various approximations in this appendix are expansions in $\ell^2$.
If these condition are fulfilled we can assume that all harmonic
oscillators are in the ground state.

\subsubsection{Semiclassical limit}
\label{sec:semiclassical}

Our first attempt to identify a collective coordinate was based
on the intuitive picture of harmonic oscillators which are not
excited for small values of $\ell$. In a second independent
approach we determine a trajectory which can be used to describe
quantum tunneling in the full system in the semiclassical limit.

A standard method to derive semiclassical expressions for
quantum tunneling are
instantons~\cite{Rajaraman-1982,Coleman-1985,Kleinert-1993,Weiss-1999}.
In this method Feynman path integrals are extended to imaginary
times. For simple systems where the Lagrangian can be written
as the difference between kinetic energy and potential energy
this is equivalent to path integrals in real time for the inverted
potential. As in the ordinary Feynman path integral formalism in
quantum mechanics, semiclassical quantum mechanics is based on
classical trajectories and small fluctuations around these
trajectories.

We use the instanton technique only to determine the classical trajectory
that governs quantum tunneling in the semiclassical limit. As in our
first approach this trajectory is then used to define a collective coordinate.
In this appendix we do not use instantons to derive semiclassical expressions
for the full system. This would require to take into account fluctuations
around this trajectory.

In our simplified approach we have to find
the classical trajectory that starts with zero kinetic energy at one
of the maxima of the inverted potential and ends with zero kinetic energy
at the other maximum. Due to the symmetry of the potential $V(Q,\{q_n\})$
it is sufficient to find the trajectory which starts at
\begin{equation}
\label{eq:semicl-minimum}
Q=0, \quad q_n=0
\end{equation}
(minimum of the inverted potential) at $\tau=0$ and reaches
\begin{equation}
\label{eq:semicl-maximum}
Q=\frac{\pi}{2}, \quad
q_n=\frac{2\sqrt{2}\,\ell^2}{\pi^3 (2n+1)^3}
\end{equation}
(maximum of the inverted potential) for $\tau\to\infty$.

If we assume that we can express $q_n$ as a function of $Q$, that is,
$q_n(\tau)=q_n(Q(\tau))$, we can rewrite the equations of motion of a
particle moving in the inverted potential $-V(Q,\{q_n\})$ in the form
\begin{subequations}
\begin{eqnarray}
\label{eq:semicl-motion-Q}
&& \ddot{Q} - \frac{\partial V}{\partial Q} = 0 ,
\\
\label{eq:semicl-motion-q}
&& \dot{Q}^2 \frac{d^2 q_n}{dQ^2}
+ \frac{\partial V}{\partial Q} \frac{dq_n}{dQ}
- \frac{\partial V}{\partial q_n} = 0 .
\end{eqnarray}
\end{subequations}
Energy conservation requires
\begin{eqnarray}
\label{eq:semicl-energy}
&& \frac{1}{2} \dot{Q}^2
\bigg[ 1 + \sum\limits_{n=0}^{\infty} \Big(\frac{dq_n}{dQ}\Big)^2 \bigg]
- V(Q,\{q_n\})
\nonumber\\
&& = - V_0(\pi/2) = -1 + \frac{\ell^2}{24}
\end{eqnarray}
which can be used to express $\dot{Q}^2$ as a function of $Q$.
Substituting this relation into Eq.~(\ref{eq:semicl-motion-q})
we obtain, together with the boundary conditions Eqs.~(\ref{eq:semicl-minimum})
and (\ref{eq:semicl-maximum}), a system of coupled nonlinear differential
equations for the trajectory we are looking for.

Instead of solving the differential equations for $q_n(Q)$ approximately
we use the results we obtained so far. These results suggest that $q_n(Q)$
is approximately given by $q_n(Q) \approx \bar{q}_n(Q)$. We verify this
by substituting
\begin{equation}
q_n(Q)  = \bar{q}_n(Q) + \ell^6 \eta_n(Q)
\end{equation}
into Eqs.~(\ref{eq:semicl-motion-q}) and (\ref{eq:semicl-energy}). We assume
that $\eta_n(Q)$ is of the order one. From Eq.~(\ref{eq:semicl-energy})
we obtain
\begin{equation}
\label{eq:semicl-Qdot}
\dot{Q}^2 = \frac{\ell^2}{12} \cos^2 Q + O(\ell^6) .
\end{equation}
Furthermore, we use
\begin{subequations}
\begin{eqnarray}
\frac{\partial V}{\partial Q} & = &
- \frac{\ell^2}{12} \sin Q \cos Q + O(\ell^{10}) ,
\\
\frac{\partial V}{\partial q_n} & = &
\pi^2 \ell^4 (2n+1)^2 \eta_n(Q) .
\end{eqnarray}
\end{subequations}
If we take into account that $\bar{q}_n(Q)$ is of the order
$\ell^2$ [see Eq.~(\ref{eq:q-bar})] the lowest-order terms of
Eq.~(\ref{eq:semicl-motion-q}) read
\begin{equation}
\frac{\ell^2}{12} \cos Q \frac{d}{dQ} \cos Q \, \frac{d \bar{q}_n}{dQ}
- \pi^2 \ell^4 (2n\!+\!1)^2 \eta_n(Q) = 0
\end{equation}
and we immediately find
\begin{equation}
\eta_n(Q) \approx - \frac{\sqrt{2}}{3\pi^5 (2n+1)^5} \sin Q \cos^2 Q .
\end{equation}

This result confirms that $\eta_n(Q)$ is indeed of the order one. We
can therefore approximate the trajectory along which we have to calculate
the Euclidian action $S_0$ by
\begin{equation}
q_n(Q) \approx \bar{q}_n(Q) .
\end{equation}

With the help of Eq.~(\ref{eq:semicl-Qdot}) and after omitting a constant
term in the potential energy we can calculate the Euclidian action along
this trajectory without knowing $Q(\tau)$ as follows:
\begin{eqnarray}
\label{eq:S_0}
\frac{S_0}{\hbar} & \approx  &\frac{E_J\lambda_J \ell}{\hbar\omega_p}
\int\limits_{-\infty}^{\infty} \left[
   \frac{1}{2} \dot{Q}^2 + \frac{\ell^2}{24}\cos^2 Q \right] d\tau
\nonumber\\
& = & \frac{E_J\lambda_J \ell}{\hbar\omega_p}
\int\limits_{-\pi/2}^{\pi/2} \frac{1}{\dot{Q}} \left[
   \frac{1}{2} \dot{Q}^2 + \frac{\ell^2}{24}\cos^2 Q\right] dQ
\nonumber\\
& = & \frac{E_J\lambda_J \ell}{\hbar\omega_p}
\int\limits_{-\pi/2}^{\pi/2} \frac{\ell}{\sqrt{12}} \cos Q \,\, dQ
\nonumber\\
& = & \frac{1}{\sqrt{3}} \frac{E_J\lambda_J}{\hbar\omega_p} \ell^2
= \frac{4 U_0}{\hbar \omega_0} .
\end{eqnarray}
Here we only have taken into account terms of the order $\ell^2$.
In the last step we have used the definitions of $U_0$ and
$\omega_0$, Eqs.~(\ref{eq:U_0}) and (\ref{eq:omega_0}).

It is important to note, that Eq.~(\ref{eq:S_0}) can be read
in two ways: According to our calculations, it describes the approximate
Euclidian action along the trajectory defined by $q_n(\tau) =
\bar{q}_n(Q(\tau))$. Obviously, it also describes the exact Euclidian action
of a particle moving in the potential $\tilde{U}(Q) = E_J \lambda_J \ell^3
\cos^2 Q / 24$. Therefore, in the semiclassical limit we obtain the correct
exponent for the energy splitting when we use $Q(\tau)$ as a collective
coordinate to approximate the full system by a system with a single
degree of freedom.

In the semiclassical limit the exponent $S_0/\hbar$ becomes large.
Taking into account that various approximations in this appendix are
expansions in $\ell^2$, the semiclassical limit is characterized by
\begin{equation}
\frac{\hbar\omega_p}{E_J\lambda_J} \ll \ell^2 \ll 1
\end{equation}
within the framework of this appendix. If the parameters of the
Josephson junction fulfill this condition, our collective coordinate
$Q(\tau)$ seems to be a useful concept to describe quantum tunneling
in short LJJs.

\subsection{Summary}

In this appendix we have shown that for appropriate parameters, the
Josephson phase $\mu(x,t)$ can approximately be described by a collective
coordinate $Q(t)$. We have expanded the Josephson phase $\mu(x,t)$ into
a Fourier series and used two approaches to reduce the many degrees of
freedom of the system to one degree of freedom.

In the first approach we used the intuitive picture of harmonic oscillators
which are not excited for small $\ell$. In the second approach we determined
the trajectory which allows us to calculate the exponent of the semiclassical
expression for the energy splitting. In lowest order in $\ell^2$ we obtained
the same collective coordinate $Q(t)$ for both approaches.

\section{Analytical results}
\label{App:Analytics}

As discussed in Sec.~\ref{Sec:0-pi-JJ} and Appendix~\ref{App:CollCoord}, our collective coordinate $Q$ describes the dynamics of a single particle of mass $M$ [see Eq.~(\ref{eq:mass})] in a one-dimensional potential $U(Q)$ [see Eq.~(\ref{eq:Ueff})]. The corresponding stationary Schr\"odinger equation reads
\begin{equation}
  \left[ - \frac{\hbar^2}{2M} \frac{d^2}{dQ^2} + U(Q) \right] \psi(Q) 
  = E \psi(Q)
  . \label{eq:schroedinger}
\end{equation}
Since we have $U(Q) \propto \cos^2 Q$ this is essentially
the well-known Mathieu equation~\cite{Abramowitz-1970}.

\subsection{Mathieu equation}
\label{App:Mathieu}

\newcommand{\ce}{\mathop{\mathrm{ce}}\nolimits}
\newcommand{\se}{\mathop{\mathrm{se}}\nolimits}

When we take into account the expressions for $M$, Eq.~(\ref{eq:mass}),
and $U(Q)$, Eq.~(\ref{eq:Ueff}), and introduce the dimensionless
parameter
\begin{equation}
  h = \left(\frac{E_J \lambda_J}{\hbar\omega_p}\right)^2\frac{\ell^4}{48}
  = \left(\frac{U_0}{\hbar\omega_0}\right)^2
  \label{eq:h-def}
\end{equation}
and the scaled energy
\begin{equation}
  \varepsilon = 2\ell
  \left(\frac{E_J \lambda_J}{\hbar\omega_p}\right)^2
  \frac{E}{E_J \lambda_J}
  = \frac{96 h}{\ell^3} \frac{E}{E_J \lambda_J}
  \label{eq:epsilon-def}
\end{equation}
we obtain the standard form
\begin{equation}
  \psi''(Q) + \left[\varepsilon -2h \cos(2Q) \right]\psi(Q) = 0
  \label{eq:mathieu}
\end{equation}
of the Mathieu equation~\cite{Abramowitz-1970}. Here we have omitted
an unimportant $\ell$-dependent term in the potential energy.

Since the Josephson phases $\mu(x,t) + 2\pi$ and $\mu(x,t)$ cannot be distinguished, $\mu(x,t)$ (and therefore also $Q(t)$) is only defined modulo $2\pi$. Therefore, we supplement Eqs.~(\ref{eq:schroedinger}) and (\ref{eq:mathieu}) with periodic boundary conditions $\psi(Q+2\pi) = \psi(Q)$.

From the definition of the parameters $h$ [see Eq.~(\ref{eq:h-def})] and $\varepsilon$ [see Eq.~(\ref{eq:epsilon-def})], we find that the quantum mechanical behavior of the system is governed by three parameters: The purely geometrical scaled length $\ell$, the energy scale $E_J \lambda_J$ of the system, and the dimensionless parameter $\hbar\omega_p/(E_J \lambda_J)$. We want to emphasize that in Eq.~(\ref{eq:mathieu}) the only parameter is $h$, since $\varepsilon$ is the eigenvalue that we want to determine.  Furthermore, $h$ does not depend on the energy scale $E_J\lambda_J$ of the system. 

For a given value of $h$, Eq.~(\ref{eq:mathieu}) has $\pi$ periodic and $2\pi$-periodic solutions only for characteristic values of $\varepsilon$. These solutions are either even or odd. Other solutions of Eq.~(\ref{eq:mathieu})
do not fulfill $\psi(Q+2\pi) = \psi(Q)$.  Following the notation of Abramowitz
and Stegun~\cite{Abramowitz-1970}, we denote the even solutions by $\ce_n(Q,h)$
and their characteristic values by $a_n(h)$ $(n=0,1,2,3, \dots)$.  Similarly,
the odd solutions and their characteristic values are denoted by $\se_n(Q,h)$
and $b_n(h)$ $(n=1,2,3, \dots)$, respectively. A closer analysis shows that
the scaled ground-state energy is given by $\varepsilon=a_0$ whereas the
scaled energy of the first excited state is given by $\varepsilon = b_1$.
The corresponding (unnormalized) eigenfunctions are $ce_0(Q,h)$ (symmetric
and $\pi$ periodic) and $se_1(Q,h)$ (antisymmetric and $2\pi$-periodic),
respectively.

\subsection{Approximate results}
\label{App:Mathieu:ApproxRes}

Approximations for the characteristic values $a_n$ and $b_n$ are available for $h \ll 1$ and $h \gg 1$, see Ref.~\onlinecite{Abramowitz-1970}. We can use these approximations to find approximations for the difference $\Delta\varepsilon = \varepsilon_2-\varepsilon_1=b_1 -a_0$ between the two lowest eigenvalues of Eq.~(\ref{eq:mathieu}).

\subsubsection{Low energy barrier $(h \ll 1)$}
\label{App:Mathieu:Approx:LowBar}

For $h \ll 1$ (low energy barrier) we find
\begin{equation}
  \Delta\varepsilon \approx 1 - h + \frac{3}{8}h^2.
  \label{eq:small-h}
\end{equation}
Using the definitions of $h$ and $\varepsilon$, Eqs.~(\ref{eq:h-def})
and (\ref{eq:epsilon-def}), we may therefore approximate the difference
$\Delta E$ between the two lowest eigenvalues of the Schr\"odinger
equation~(\ref{eq:schroedinger}) by
\begin{eqnarray}
  \Delta E & \approx & \frac{E_J \lambda_J}{2\ell}
   \left(\frac{\hbar\omega_p}{E_J\lambda_J}\right)^2
   \left[1 
     - \frac{1}{48} \left(\frac{E_J\lambda_J}{\hbar\omega_p}\right)^2 \ell^4
         \right.\nonumber\\&&\left.\hspace*{5em}
     + \frac{1}{6144}\left(\frac{E_J\lambda_J}{\hbar\omega_p}\right)^4\ell^8
   \right]
\end{eqnarray}
if $\ell$ satisfies
\begin{equation}
  \ell^4 \ll 48 \left(\frac{\hbar\omega_p}{E_J\lambda_J}\right)^2.
\end{equation}
The lowest-order term corresponds to $h=0$, that is, no energy barrier.

\subsubsection{High energy barrier $(h \gg 1)$}
\label{App:Mathieu:Approx:HighBar}

For $h \gg 1$ (high energy barrier) we find the exponentially small
difference
\begin{equation}
  \Delta \varepsilon \approx 
  32 \sqrt{2/\pi} h^{3/4} e^{-4\sqrt{h}}
  . \label{eq:large-h}
\end{equation}
Substituting the expressions for $h$ [Eq.~(\ref{eq:h-def})], $U_0$
[Eq.~(\ref{eq:U_0})], and $\omega_0$ [Eq.~(\ref{eq:omega_0})], we obtain
from Eq.~(\ref{eq:large-h})
\begin{equation}
  \Delta E \approx  8\, \hbar\omega_0
  \sqrt{\frac{2 U_0}{\pi\hbar\omega_0}} \,
  \exp \left[ - 4 \, \frac{U_0}{\hbar\omega_0} \right]
  \label{eq:delta-periodic}
\end{equation}
as the the semiclassical expression for the energy level splitting for a particle in a periodic $\cos(2Q)$ potential. Note, that the exponent
is the Euclidian action $S_0/\hbar$ [see Eq.~(\ref{eq:S_0})] that we
calculated in Sec.~\ref{sec:semiclassical}.

When we compare this result to the well-known expression
\begin{equation}
\Delta E_{\rm quartic} \approx  8\, \hbar\omega_0
\sqrt{\frac{2 U_0}{\pi\hbar\omega_0}} \,
\exp \left[ - \frac{16}{3} \,
\frac{U_0}{\hbar\omega_0} \right]
\label{eq:delta-quartic}
\end{equation}
for a particle in a quartic double-well potential (see for example
Ref.~\onlinecite{Kleinert-1993,Weiss-1999}), we immediately see, that these two
expressions only differ by a numerical factor of the order one ($4$ vs.\ $16/3$)
in the exponent. This confirms that the asymptotic expression
(\ref{eq:delta-periodic}) is indeed the semiclassical limit for
the energy level splitting. Note, that due to our periodic boundary
conditions, ``tunneling to the left'' leads to the same final state
as ``tunneling to the right''. We should therefore actually compare
$\Delta E$ to $2\Delta E_{\rm quartic}$.

\subsection{Two-level system}
\label{App:Mathieu:2level}

In the present paper we use the energy difference between the two lowest eigenvalues of the Schr\"odinger equation to describe coherent quantum oscillations between the two classical stationary states of the system. This simple approach is only valid as long as the energy barrier is sufficiently high. In this case the difference between the two lowest energy eigenvalues (energy splitting) is small. Furthermore, the sum and the difference between the eigenfunctions corresponding to the two lowest energy eigenvalues are well localized in one of the minima of the potential $U(Q)$ and the two lowest energy levels are sufficient to describe coherent quantum oscillations.

If the barrier becomes too small we need more energy levels to describe the quantum dynamics of the system sufficiently well.  Therefore, we require that the difference between the two lowest energy eigenvalues is significantly smaller than the energy difference between the first excited state and the higher energy eigenstates.  Numerical calculations based on the \textsc{Matematica} functions \texttt{MathieuCharacteristicA} and \texttt{MathieuCharacteristicB} show that for $h\approx 0.348950$ the difference $\Delta\varepsilon_{12} = \varepsilon_2-\varepsilon_1$ between the two lowest eigenvalues is the same as the energy difference $\Delta\varepsilon_{23} = \varepsilon_3-\varepsilon_2$ between the first and the second excited state and is given by $\Delta\varepsilon_{12} = \Delta\varepsilon_{23} \approx 0.696576$.  Therefore, we have to choose the parameters of our system such that $h$ is significantly larger than $0.35$. A more detailed analysis shows that $\Delta\varepsilon_{23}$ exceeds $\Delta\varepsilon_{12}$ by a factor of 2 for $h > 0.558721$, by a factor of 5 for $h > 0.936175$, and by a factor of 10 for $h > 1.293984$.

The conditon that $h$ is significantly larger than $0.35$ requires that $\ell^4$ is significantly larger than $16.8 (\hbar\omega_p/E_J\lambda_J)^2$. As a consequence, we are always in a regime where we cannot approximate the system by a two-level system if the approximation~(\ref{eq:small-h}) is applicable.

Finally we want to mention, that for $h\approx 0.329006$ the energy of the first excited state touches the energy barrier. For smaller values of $h$ only the lowest energy level is below the energy barrier.

\bibliography{MyJJ,jj,jj-annular,LJJ,SFS,pi,SF,software,QuComp}

\end{document}